\documentclass[preprint]{aastex}

\shorttitle{Multiband Optical Observation of P/2010 A2 Dust Tail}
\shortauthors{Kim et al.}

\begin{document}

\title{Multiband Optical Observation of P/2010 A2 Dust Tail}

\author{Junhan Kim\footnote{Graduated from Seoul National University,
Republic of Korea in August 2010}} 
\affil{3\#\#-\#\#\#, Mokdong Apartment, Mok-5-dong, Yangcheon-gu, Seoul
158-753, Republic of Korea}

\author{Masateru Ishiguro\footnote{Corresponding author. E-mail :
ishiguro@astro.snu.ac.kr}} 
\affil{Department of Physics and Astronomy, Seoul National University,
San 56-1, Silim-dong, Gwanak-gu, Seoul 151-742, Republic of Korea} 

\author{Hidekazu Hanayama}
\affil{Ishigakijima Astronomical Observatory, National Astronomical
Observatory of Japan, Ishigaki, Okinawa 907-0024, Japan} 

\author{Sunao Hasegawa and Fumihiko Usui}
\affil{Institute of Space and Astronautical Science, Japan Aerospace
Exploration Agency, 3-1-1 Yoshinodai, Chuo-ku, Sagamihara, Kanagawa
252-5210, Japan} 

\author{Kenshi Yanagisawa}
\affil{Okayama Astrophysical Observatory, National Astronomical
Observatory of Japan, Asaguchi, Okayama 719-0232, Japan} 

\author{Yuki Sarugaku}
\affil{Institute of Space and Astronautical Science, Japan Aerospace
Exploration Agency, 3-1-1 Yoshinodai, Chuo-ku, Sagamihara, Kanagawa
252-5210, Japan} 

\author{Jun-ichi Watanabe}
\affil{National Astronomical Observatory of Japan, Mitaka, Tokyo, 181-8588, Japan}

\and

\author{Michitoshi Yoshida}
\affil{Hiroshima Astrophysical Science Center, Hiroshima University, 
Higashi-Hiroshima, Hiroshima 739-8526, Japan}

\begin{abstract}
An inner main-belt asteroid, P/2010 A2, was discovered on January 6th, 2010.
Based on its orbital elements, it is considered that the asteroid belongs to the Flora collisional family, 
where S-type asteroids are common, whilst showing a comet-like dust tail.
Although analysis of images taken by the Hubble Space Telescope and Rosetta spacecraft suggested that the dust tail resulted 
from a recent head-on collision between asteroids \citep{jewitt10,snod10}, 
an alternative idea of ice sublimation was suggested based on the morphological fitting of ground-based images \citep{mor10}.
Here, we report a multiband observation of P/2010 A2 made on January 2010 with a 105 cm telescope at the Ishigakijima Astronomical Observatory.
Three broadband filters, $g'$, $R_c$, and $I_c$, were employed for the observation.
The unique multiband data reveals that the reflectance spectrum of the P/2010 A2 dust tail resembles that of an Sq-type asteroid or 
that of ordinary chondrites rather than that of an S-type asteroid.
Due to the large error of the measurement, the reflectance spectrum also resembles the spectra of C-type asteroids, even though C-type asteroids are uncommon in the Flora family.
The reflectances relative to the $g'$-band (470 nm) are 1.096$\pm$0.046 at the $R_c$-band (650 nm) and 1.131$\pm$0.061 at the 
$I_c$-band (800 nm).
We hypothesize that the parent body of P/2010 A2 was originally S-type but was then shattered upon collision into scaterring fresh chondritic particles 
from the interior, thus forming the dust tail.
\end{abstract}


\section{Introduction}

Main-belt asteroid P/2010 A2 (LINEAR, hereafter P/2010 A2) was discovered on January 6th, 2010 \citep{birt10} 
by Lincoln Near-Earth Asteroid Research (LINEAR).
It showed a comet-like dust tail without a central coma, representing a remarkable distinction from ordinary
comets \citep{mor10,jewitt10,snod10,hain11}.
The mechanism of the comet-like activity remains under debate.
\citet{mor10} applied inverse dust tail Monte Carlo fitting to their observational images and 
argued that water-ice sublimation brought about the cometary activity.
However, three independent research groups stated that collisional disruption
was responsible for the dust tail on the basis of their observations and dynamical studies \citep{jewitt10,snod10,hain11}.
Note that more than 10 months had already passed since the dust emission.
Details such as impact ejecta plume and jet structure may not have been captured in their observed images \citep{hsieh11,bode11,ishi11}. 

The proper orbital elements of P/2010 A2 are a semi-major axis $a=2.29$ AU,
an eccentricity $e=0.12$, and an inclination $i=5.26^{\circ}$.
It belongs to a dynamical asteroid group, the Flora family, whose proper orbital
elements are $2.12<a<2.31$AU, $0.11<e<0.175$, and $3^{\circ} < i < 7.5^{\circ}$ \citep{nes02}.
It is important to note that P/2010 A2 remains within the snowline of the Solar System. 
The orbital aspect differentiates it from known main-belt comets, 
which have an aphelion greater than 2.5 AU \citep{hsieh06,hsieh09,jewitt11}. 
In addition to their dynamical properties, asteroids are classified in accordance
with the shape of their spectra \citep[e.g.,][]{bus02_ast}.
Furthermore, asteroid families share similar spectral properties, and therefore, family members belong to the
similar taxonomic types \citep{cel02}.
Most of the Flora members are categorized as S-type asteroids \citep{flo98}.
However, S-type asteroids are known to be parent bodies of ordinary chondrites, not containing obvious evidence for water ice,
unlike C-type or D-type asteroids \citep{camp10,lica11}. 

To date, the taxonomic type of P/2010 A2 has not been reported in refereed journals. 
In this paper, we present the photometric observation of P/2010 A2 made at the Ishigakijima Astronomical Observatory,
Japan, by simultaneous imaging with $g'$, $R_c$, and $I_c$-band filters.
We describe the data reduction and calibration process used to derive the reflectance spectrum
of the object from the integrated fluxes at different wavelengths.
By comparing the reflectance to the spectra of different types of known asteroids,
we discuss the cause of the comet-like activity of P/2010 A2.

\section{Observation and Data Reduction}

All P/2010 A2 images used in this research were obtained on 15th January, 2010 (15:56 -- 20:02 UT) with a 105 cm telescope of Ishigakijima Astronomical Observatory (IAO), National Astronomical Observatory of Japan (NAOJ),
located on Ishigaki Island, Okinawa, Japan. 
On the observation date, the object's heliocentric distance was 2.014 AU, geocentric distance was 1.043 AU, and solar phase angle was $6.45^{\circ}$.
The telescope at IAO, named `{\it Murikabushi}' has a 105 cm diameter. 
The optical system is a Ritchey-Chr\'etien and the focal length is 1260 cm (F12). 
At the Cassegrain focal plane, a three-channel simultaneous imaging system with
three CCD cameras (MITSuME: Multicolor Imaging Telescopes for Survey and Monstrous Explosions) was attached for the observation.
This system has band-pass filters of SDSS (Sloan Digital Sky Survey) $g'$, Johnson--Cousins $R_c$, and $I_c$.
Each CCD camera consists of an Alta U6 (Apogee Instruments Inc.)
with an array size of $1024 \times 1024$ pixels and a pixel size of $24 \times 24~{\rm \mu m}$.
We took 44 images for each wavelength under clear conditions: 24 images were
taken with an exposure time of 180 s, and a further 20 images were attained with an exposure time of 300 s.
In total, we have an integration time of 172 minutes.

Concerning the data reduction, we focused on subtracting background stars from all the images to obtain a faint dust tail \citep{ishi07, ishi08}.
This was accomplished using our own programs to remove or mask stars and also using well-known software such as {\it IRAF}, {\it SExtractor}
\citep{bert96}, and {\it WCStools} \citep{mink97}.
Firstly, we prepared star-aligned composite images using WCS (World Coordinate System) information
in order to detect background sources including stars and galaxies.
WCS information was necessary to precisely determine the locations of the object in every frame.
Then, bright sources other than the object were automatically found by {\it SExtractor}.
A star-aligned image is more useful than a single image because it enables easier detection of faint stars. 
Then, circular masks, with zero-values and sizes corresponding to the width of the point spread function of individual stars,
were applied to the location of stars (see Figure 1).
Finally, combining all the masked frames with offsets to align
the images with respect to the position of P/2010 A2 gave us an image showing only a faint dust tail.
In order to calculate the offsets, the position of the object was computed at a given time using our program,
and the $sky2xy$ command in $WCStools$ was utilized to convert the equatorial coordinate to a CCD image coordinate.
The masked images were combined with offsets to give the image of the dust tail,
excluding the masked pixels and shifting the background intensity to a zero level.
Since P/2010 A2 moved relative to the background objects, it is possible to exclude the masked data completely,
and to combine the sequence of images.
Finally, we obtained an image without stars and galaxies. These resulting images are shown in Figure 2.

\section{Analysis}

\subsection{Flux Calibration}

To integrate a flux coming from the object for each wavelength,
a cut profile along the perpendicular direction of the dust tail was analyzed.
Assuming an identical flux distribution for different wavelength bands,
a region was selected(see Figure 3(a)) to extract the profile.
For precise measurements on this extended structure, the flux was averaged along the line of the tail.
This was done using the $projection$ function in on {\it SAOImage ds9} \citep{joye03}.
The background flux coming from the sky brightness was fitted to a polynomial
function and subsequently subtracted from the original profile. 

Figure 3(b) shows cut profiles for three wavelengths.
Note that a raw intensity value (ADU, arbitrary data unit) was used here to present the flux. 
This raw intensity value can be scaled to other physical units using conversion parameters determined from standardization, as shown below.

\subsection{Reflectance} 

In order to convert the ratio of flux for three wavelengths into relative reflectances,
we need photometric transformation equations as well as the color indices of the Sun.
The transformation equations were derived by photometry from standard star data.
Unfortunately, our observation of the standard field contained several stars whose standard magnitudes at $g'$-band are unknown.
Based on the magnitudes of standard stars listed in \citet{land92},
we found the UBV${R_c}{I_c}$ photometric magnitudes and converted them to Sloan $u'g'r'i'z'$ magnitudes.
Conversion equations between the two systems are given by \citet{smith02}.
After the arithmetic operations, a standard star chart of the $g' {R_c} {I_c}$ system was prepared for analysis,
and then, the parameters for the photometric transformation equations were obtained using the $fitparams$ function in{\it IRAF}.

The solar color indices for various filter systems are well known.
However, the filter system used in our observation is a combination of two photometric systems :
$g'$ of the SDSS system, $R_c$ and $I_c$ of the Cousins system.
Therefore, solar color indices such as $(g'-R_c)_{\odot}$ and $(g'-I_c)_{\odot}$ are required.
Again, transformation equations between the Johnson-Cousin (UBV$R_{c}I_{c}$) system and the SDSS ($u'g'r'i'z'$) system of \citet{smith02} were used :

\begin{equation}
V = g' - 0.55(g' - r') - 0.03
\end{equation}
\begin{equation}
V - {R_{c}} = 0.59(g' - r') + 0.11
\end{equation}
\begin{equation}
{R_{c}} - {I_{c}} = 1.00(r' - i') + 0.21
\end{equation}

Combining the above equations and applying the solar indices of $(g'-r')_{\odot} = 0.45$ and $(r'-i')_{\odot} = 0.10$ \citep{ivez01},
we obtain $(g'-R_c)_{\odot} = 0.653$ and $(R_c-I_c)_{\odot} = 0.963$

On the basis of these indices and transformation equations, we determined the relative reflectances compared to the reflectance at a 
particular wavelength (here, at $g'$-band : ${I_{g'}}/{I_{{\odot},g'}}$) as follows:

\begin{equation}
\left(\frac{I_{R_c}}{I_{{\odot},R_{c}}}\right) \slash \left(\frac{I_{g'}}{I_{{\odot},g'}}\right) = 10^{0.4 \times \left[(g'-R_{c})-(g'-R_{c})_{\odot} \right]}
\end{equation}
\begin{equation}
\left(\frac{I_{I_c}}{I_{{\odot},I_{c}}}\right) \slash \left(\frac{I_{g'}}{I_{{\odot},g'}}\right) = 10^{0.4 \times \left[(g'-I_{c})-(g'-I_{c})_{\odot} \right]}
\end{equation}

Note that the characters shown on the left hand side of the equations denote intensities,
which are the integrated quantities of the flux profile in Figure 3(b).
We had to exercise caution in determining the reflectance, given as an exponential function with base 10,
because it is highly sensitive to color correction and transformation of the standard photometric system.

\section{Results \& Discussion}

Table 1 shows the reflectance spectrum normalized at $g'$-band (470 nm).
In order to determine the taxonomic type of the object, we calculated the average spectra of different
taxonomic type asteroids from the Flora family using SMASS II (Phase II Small Main-Belt Asteroid Spectroscopic Survey) data \citep{bus02}.
In the archive, there are 66 asteroids in the Flora asteroid family whose spectral types are known.
The members are selected according to the asteroid dynamical family classifications \citep{zapp95}.
Among them, we found that there are 61 S-type asteroids, including 4 Sq-type asteroids
(1324 Knysna, 2873 Binzel, 2902 Westerlund, and 4733 ORO) and 2 C-type asteroids (2952 Lilliputia and 4396 Gressmann).
In Figure 4(a)--(c), the average reflectance spectra of S-type, C-type,
and Sq-type asteroids with one standard deviation are compared with that of P/2010 A2.

Firstly, we noticed that the reflectance spectrum of the P/2010 A2 dust tail is significantly
different from that of S-type asteroids (Figure 4(a)).
In Figure 4(b), we compare the P/2010 A2 spectrum to the reflectances of C-type asteroids.
A C-type asteroid has a relatively flat spectrum for $\lambda$ $>$ 550 nm but exhibits absorption for $\lambda$ $<$ 400 nm.
Although the P/2010 A2 spectrum is slightly redder than that of C-type asteroids,
we cannot rule out the possibility of it being a C-type asteroid due to large uncertainties in our measurement.
Figure 4(c) and Figure 4(d) show the reflectance spectra of Sq-type (of SMASS II class) asteroids and that of ordinary chondrites,
respectively. We obtained the spectra of ordinary chondrites from \citet{gaff01}.
These are considered as analogs of S-type asteroids but show differences in slope for $\lambda$ $<$ 700 nm and in depth of absorption
at $\lambda$ $\sim$ 950 nm. S-type asteroids show a redder spectral slope at visible wavelength,
and the depth of the absorption band at $\sim$950 nm is shallower than that of ordinary chondrites and
Sq-asteroids\citep{CHAP2004,sasaki01,ishi07b}.
Ordinary chondritic particles were also found on an S-type asteroid Itokawa\citep{NOG2011}.
It is estimated that our result concerning P/2010 A2 is similar to those of Sq-type asteroids and ordinary chondrites.
In particular, our spectrum has the closest spectral property to that of H5 chondrite.

The significance in our result is that the P/2010 A2 debris shows optical properties that are different from those of S-type asteroids.
It would have been reasonable to suppose that P/2010 A2 has a spectrum similar to that of S-type asteroids,
as they are the dominant taxonomic asteroids in the Flora family.
In fact, 92 \% (61 out of 66 asteroids) of the Flora asteroids in SMASS II are classified as either S-type or Sq-type (a subclass of S-type).
Why is P/2010 A2 different from S-type? 
As we mentioned above, there is a possibility of it being a carbonaceous asteroid,
which could contain hydrated minerals and possibly water ice.
This seems reasonable as some of the main-belt comets (MBCs), which were likely to be activated by sublimation of water and ice,
show similar spectra to those of C-type asteroids \citep{hsieh09b}.

Whereas we dismiss the possibility of P/2010 A2 being C-type,
we suggest that its debris particles have a composition similar to that of ordinary chondrites.
It would be unreasonable to assume that there was water ice in the warm region of the inner main-belt \citep{hain11}.
As the size of the parent body was small, with a diameter of 120 m \citep{jewitt10,snod10},
icy particles should have been sublimated if had they existed inside the parent bodies.
In addition, dark primitive asteroids are rare in the vicinity of P/2010 A2.
Furthermore, the inconsistency with S-type asteroids seems to be feasible even though they are the major constituents of the Flora family.
If dust particles were generated by a catastrophic collisional process, 
a large portion of the materials constituting the dust tail should originate from inside the parent body.
Such materials would be fresher than materials on the surface of S-type asteroids due to the effect of space weathering.
As pointed out by \citet{jewitt10,snod10,hain11}, we support the hypothesis of a catastrophic collision.
Our results of determining spectral type of the dust tail may be weak in the sense that they are based on only three spectral points in the visible range that prevent a definite classification of the parent body or the debris composition.
Further observations at longer wavelength (around 0.9 -- 1.0 \micron) are essential to distinguish Sq-type asteroids and C-type asteroids.

\section{Summary}

A main-belt asteroid P/2010 A2 was discovered with an unusual comet-like dust tail. We performed photometric observations of this object, using a 1 m telescope at the Ishigakijima Astronomical Observatory, to determine the reflectance properties in order to assign a spectral type.

1. For three different wavelengths, the optical images were combined after masking bright sources other than the object. Integrated fluxes coming from the object on the composite image gave the following reflectances relative to $g'$-band (470 nm): 1.096$\pm$0.046 at $R_c$-band (650 nm) and 1.131$\pm$0.061 at $I_c$-band (800 nm).

2. The proper orbital elements, namely, the semimajor axis, the eccentricity, and the inclination, led to the object being classified as part of the  Flora asteroid family, where S-type asteroids are common (estimated to account for 92\% of the Flora asteroid family, from the SMASS II archive). However, the reflectance spectrum of the P/2010 A2 dust tail is dissimilar to those of S-type asteroids.

3. Comparing the reflectance spectrum of P/2010 A2 to those of three different taxonomic types from the Flora asteroid family shows that there is a possibility of it being a C-type or an Sq-type asteroid. In addition, the spectrum resembles that of ordinary chondrites, which are considered as analogs of S-type asteroids.

4. We suggest that the dust tail of P/2010 A2 has a property similar to ordinary chondrites and that the parent body could be an S-type asteroid. If the dust particles were generated by the collision, it may be considered that fresh materials from the parent body were scattered and showed a similar  spectra to that of ordinary chondrites that can be found on an S-type asteroid.

\acknowledgments
This research was supported by the National Research Foundation of Korea. S.H. is supported by the Space Plasma Laboratory, ISAS, JAXA.

\clearpage

\begin{figure}
	\begin{center}
	$\begin{array}{cc}
		\includegraphics[width=8cm]{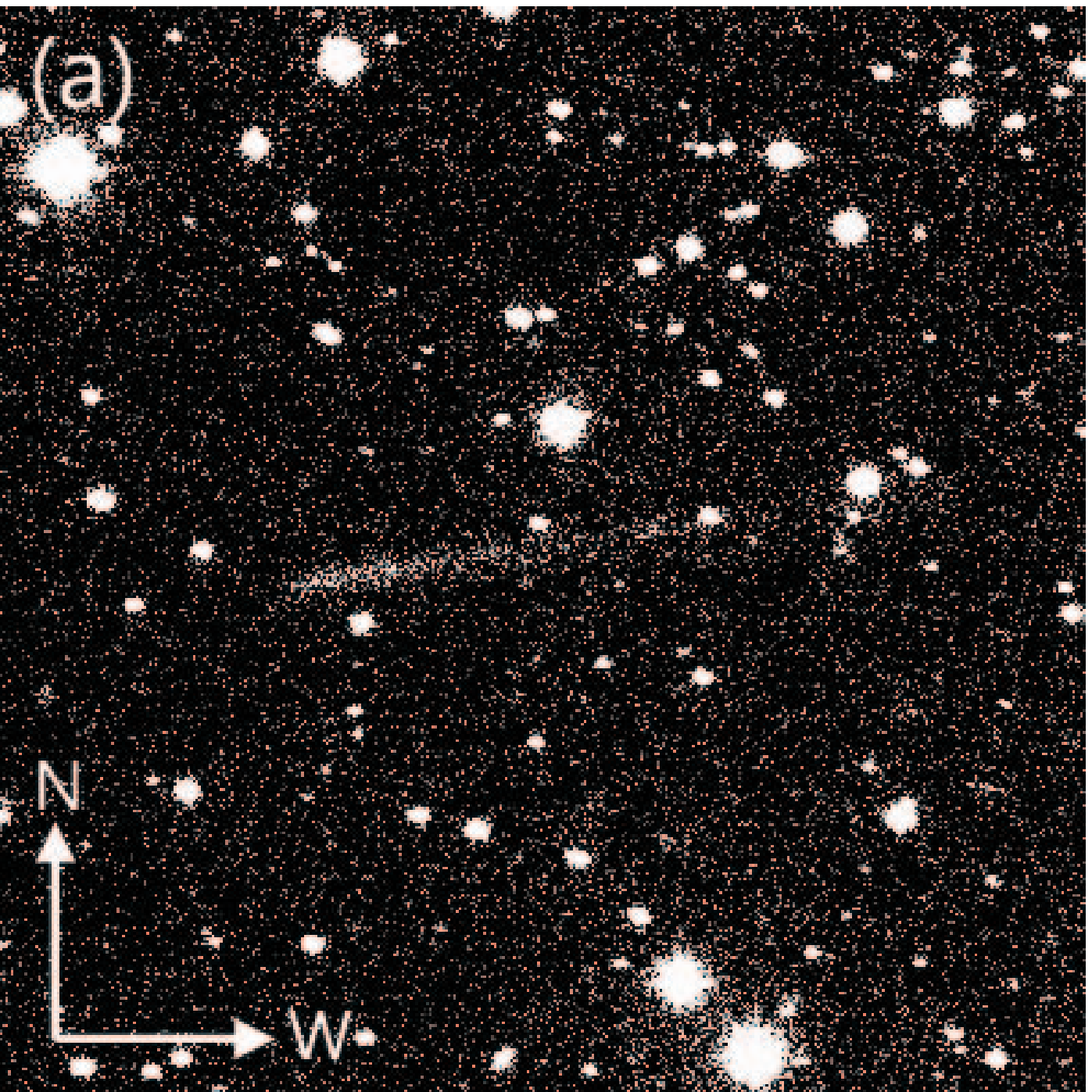} & \includegraphics[width=8cm]{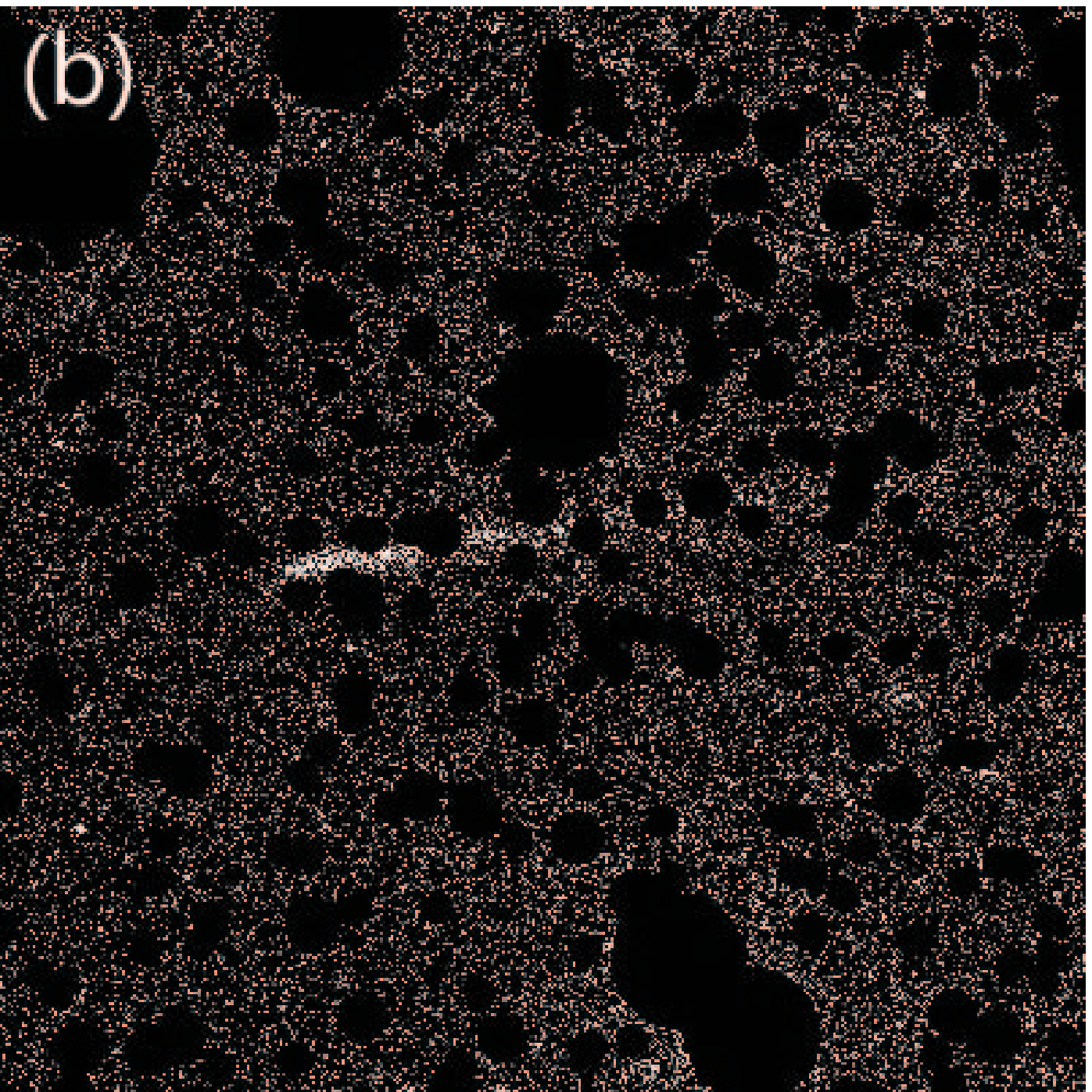} \\
	\end{array}$
	\end{center}
\caption{Masking of bright sources on the image. (a) Original image. (b) Resultant image after removing background stars. Circular masks with zero-values whose sizes were taken in accordance with brightness of each stars were put to the location of stars.}
\end{figure}

\clearpage

\begin{figure}
\begin{center}
	$\begin{array}{cc}
		\includegraphics[width=10cm]{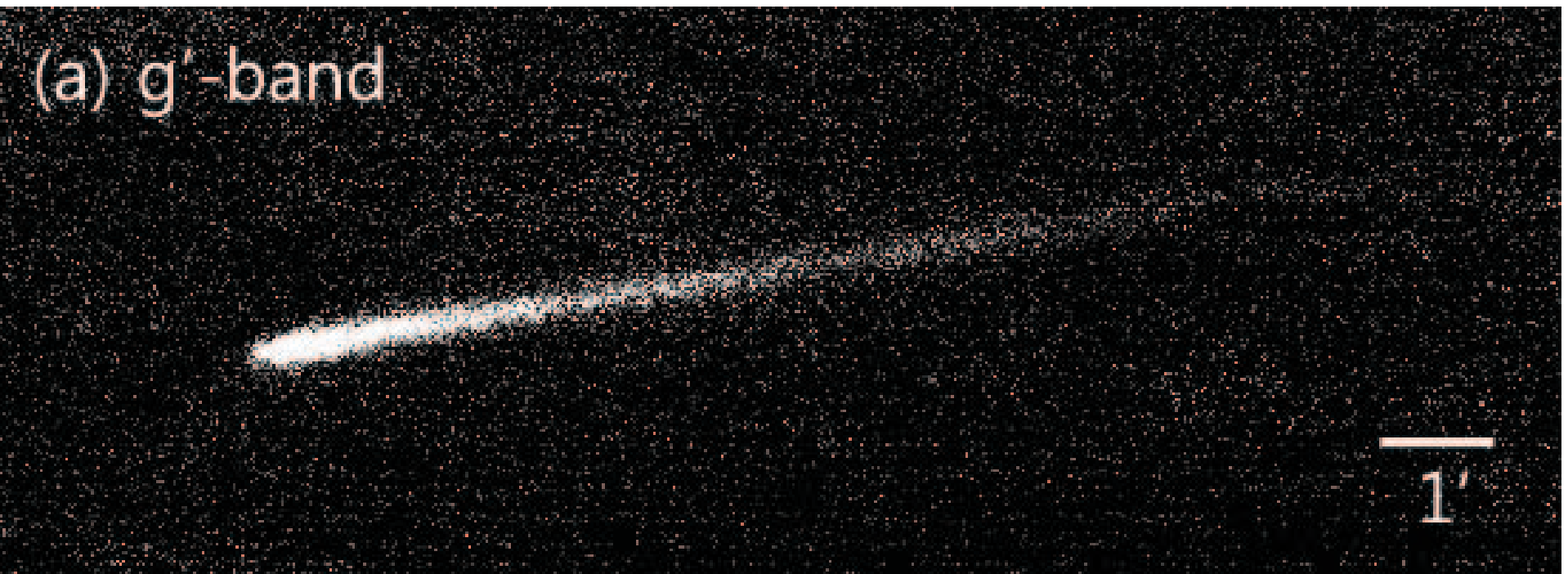}\\
		\includegraphics[width=10cm]{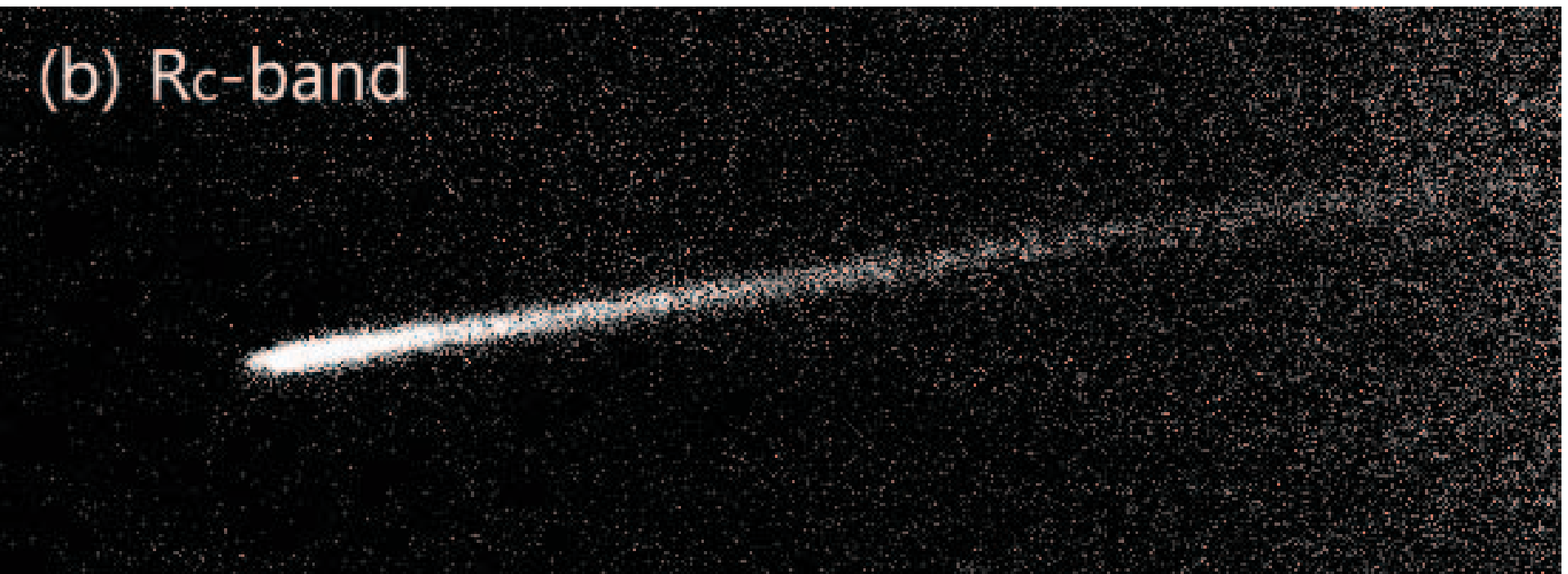}\\
		\includegraphics[width=10cm]{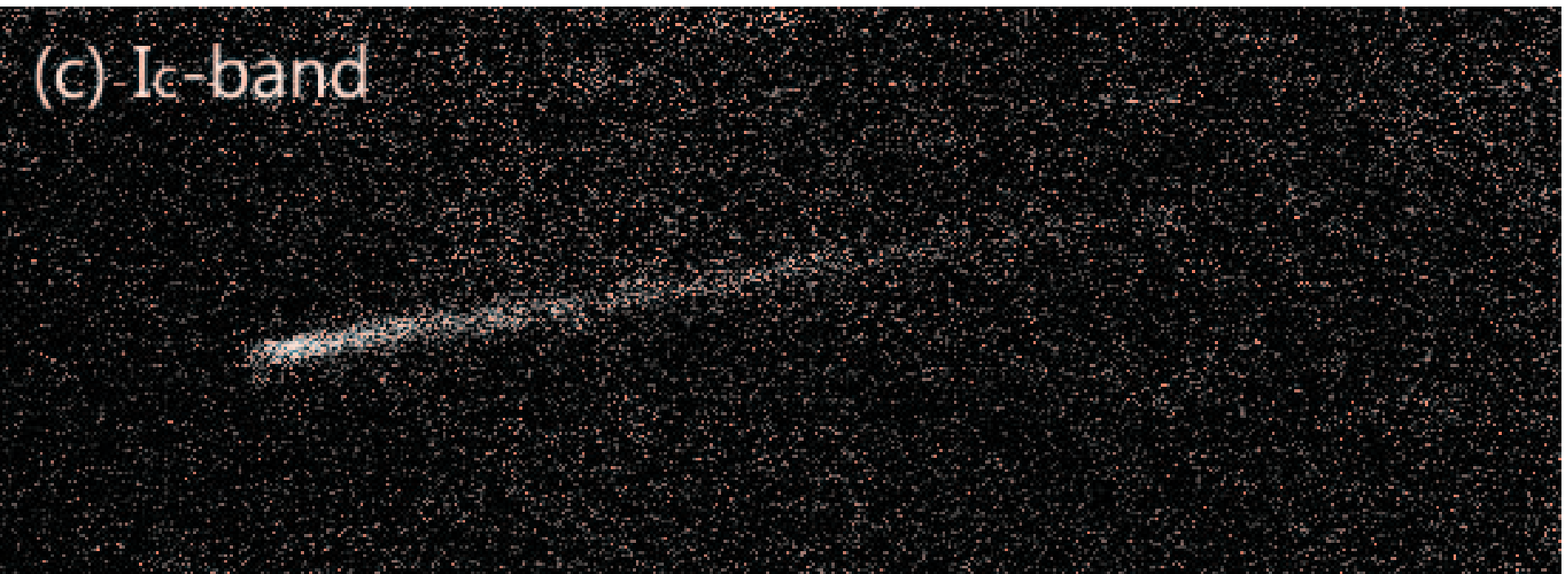}\\
	\end{array}$
	\end{center}
\caption{Images of dust tail. These are composite images of masked frames in three different waveband. P/2010 A2 at (a) $g'$-band, (b) $R_c$-band, (c) $I_c$-band}.
\end{figure}

\clearpage

\begin{figure}
\begin{center}
\includegraphics[width=10cm]{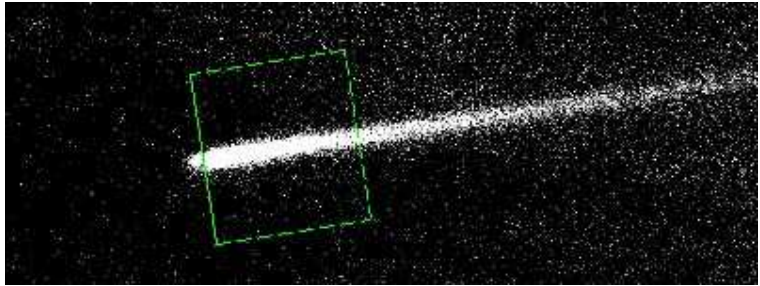} \\ \mbox{(a)}\\
\includegraphics[width=14cm]{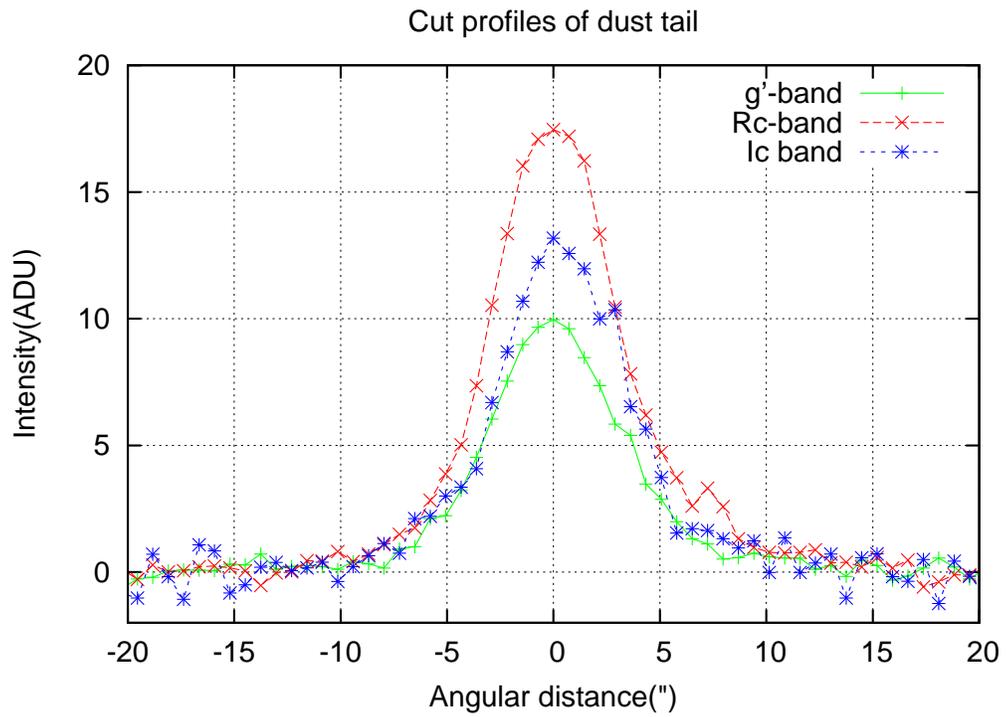} \\ \mbox{(b)}\\
\end{center}
\caption{(a) Region around the dust tail of P/2010 A2 where the flux profile is examined. (b) Cut profile of the dust tail showing averaged flux distribution perpendicular to the line of the dust tail. Relative reflectance can be obtained by comparing total fluxes and solar indices.}
\end{figure}

\clearpage

\begin{figure}
	\begin{center}
	$\begin{array}{cc}
		\includegraphics[width=8.5cm]{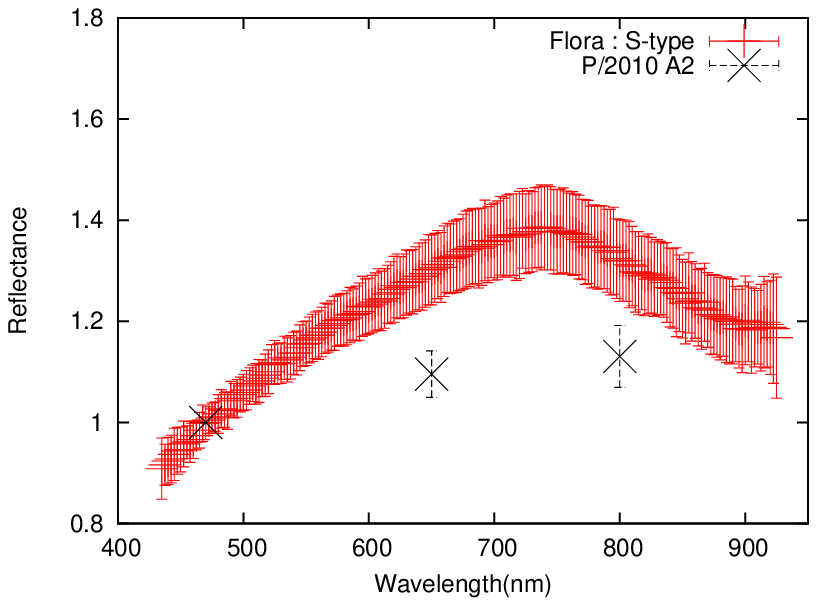} & \includegraphics[width=8.5cm]{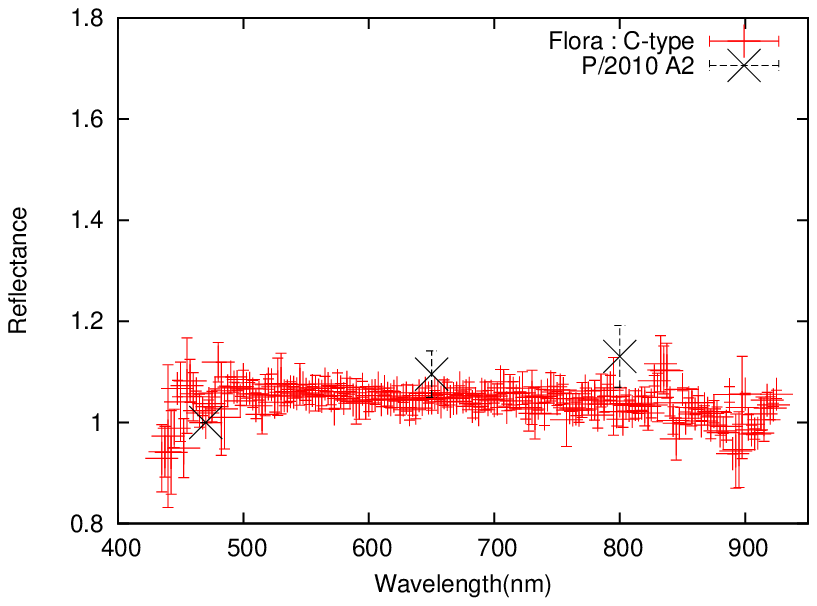} \\
		\mbox{(a)} & \mbox{(b)}\\
		\includegraphics[width=8.5cm]{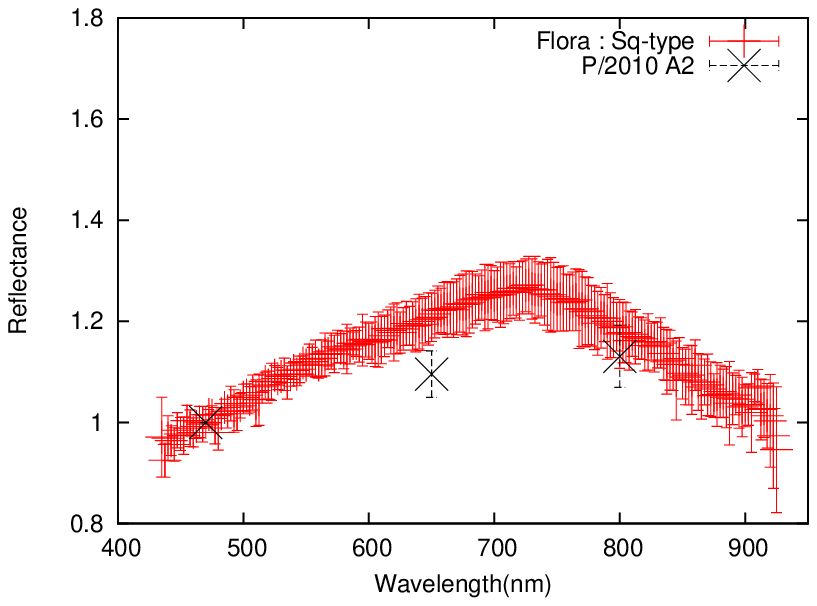} & \includegraphics[width=8.5cm]{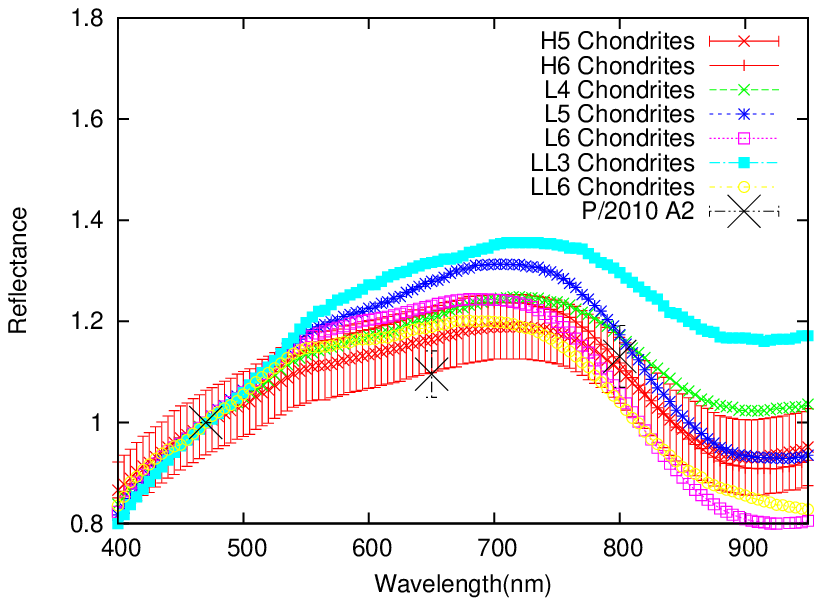} \\
		\mbox{(c)} & \mbox{(d)}\\
	\end{array}$
	\end{center}
\caption{Reflectances of P/2010 A2 and asteroids representing typical taxonomic types of Flora family members.
Reflectances are normalized at $g'$-band (470nm) here.
(a) S-type asteroids. (b) C-type asteroids. (c) Sq-type asteroids.
For each taxonomic type, average spectra with standard deviation are demonstrated and they are observation results of SMASS II, 
Phase II Small Main-Belt Asteroid Spectroscopic Survey \protect\citep{bus02}.
 (d) Ordinary chondrites. Note that error bar was indicated for reflectance of H5 chondrites, which shows the most similar spectral property to that of P/2010 A2.}
\end{figure}

\clearpage

\begin{table}
\begin{center}
\caption{Relative reflectances of P/2010 A2}
\begin{tabular}{crrr}
\tableline\tableline
Filter band & Center wavelength(nm) &Relative reflectance\tablenotemark{a} & Error\tablenotemark{b} \\
\tableline
$g'$ & 470 & 1.000 & - \\
$R_{c}$ & 650 & 1.096 & 0.046 \\
$I_{c}$ & 800 & 1.131 & 0.061 \\
\tableline
\end{tabular}
\tablenotetext{a}{Reflectances are normalized at $g'$-band.}
\tablenotetext{b}{Error of $g'$-band reflectance was regarded as zero, because we calculated relative reflectances directly from color indices, $(g'-R_{c})$ and $(g'-I_{c})$.}
\end{center}
\end{table}

\end{document}